\documentclass[12pt,a4paper]{article}
\usepackage{epsfig}
\usepackage{axodraw}
\def\nl{\nonumber\\}
\def\beq{\begin{equation}}
\def\eeq{\end{equation}}
\def\beqar{\begin{eqnarray}}
\def\eeqar{\end{eqnarray}}
\def\bfi{\begin{figure}}
\def\efi{\end{figure}}
\def\btab{\begin{table}}
\def\etab{\end{table}}
\def\bce{\begin{center}}
\def\ece{\end{center}}
\def\bit{\begin{itemize}}
\def\eit{\end{itemize}}

\def\scrs{\scriptscriptstyle}
\def\text{\textstyle}

\def\al{\alpha}

\def\de{\delta}

\def\la{\lambda}
\def\si{\sigma}

\def\refeq#1{\mbox{(\ref{#1})}}
\def\reffi#1{\mbox{Fig.~\ref{#1}}}

\def\refse#1{\mbox{Sect.~\ref{#1}}}

\def\citere#1{\mbox{Ref.~\cite{#1}}}
\def\citeres#1{\mbox{Refs.~\cite{#1}}}

\newcommand{\GeV}{\unskip\,\mathrm{GeV}}

\newcommand{\TeV}{\unskip\,\mathrm{TeV}}
\newcommand{\fba}{\unskip\,\mathrm{fb}}

\def\mathswitchr#1{\relax\ifmmode{\mathrm{#1}}\else$\mathrm{#1}$\fi}

\newcommand{\PW}{\mathswitchr W}

\def\mathswitch#1{\relax\ifmmode#1\else$#1$\fi}

\newcommand{\MW}{\mathswitch {M_\PW}}

\newcommand{\sw}{\mathswitch {s_{\scrs\PW}}}
\newcommand{\cw}{\mathswitch {c_{\scrs\PW}}}
\newcommand{\rw}{{\mathrm{W}}}

\newcommand{\cew}{C^{\ew}}

\def\ie{i.e.\ }
\def\eg{e.g.\ }

\def\wrt{wrt.\ }

\newcommand{\ord}{{\cal O}}

\newcommand{\ew}{\mathrm{ew}}

\newcommand{\ri}{\mathrm{i}}

\newcommand{\rR}{\mathrm{R}}
\newcommand{\rL}{\mathrm{L}}
\newcommand{\rT}{{\mathrm{T}}}
\newcommand{\rS}{{\mathrm{S}}}
\newcommand{\rd}{{\mathrm{d}}}
\newcommand{\rc}{{\mathrm{c}}}

\newcommand{\pT}{p_{\mathrm{T}}}
\newcommand{\pTcut}{p_{\mathrm{T}}^{\mathrm{cut}}}

\newcommand{\M}{{\cal {M}}}
\newcommand{\calL}{{\cal L}}

\newcommand{\NNLLa}{\stackrel{\mathrm{NNLL}}{=}}
\newcommand{\NLLa}{\stackrel{\mathrm{NLL}}{=}}
\newcommand{\NNLL}{\mathrm{NNLL}}
\newcommand{\NLL}{\mathrm{NLL}}

\newcommand{\LO}{\mathrm{LO}}
\newcommand{\NLO}{\mathrm{NLO}}
\newcommand{\NNLO}{\mathrm{NNLO}}
\newcommand{\shat}{{\hat s}}
\newcommand{\that}{{\hat t}}
\newcommand{\uhat}{{\hat u}}
\newcommand{\rhat}{{\hat r}}

\newcommand{\rar}{{\rightarrow}}

\newcommand{\MSBAR}{\overline{\mathrm{MS}}}
\newcommand{\msbar}{$\MSBAR$}
\newcommand{\qbar}{{\bar q}}

\def\draftdate{\relax}
\def\mpar#1{\relax}
\def\mua{\relax}
\def\mda{\relax}
\def\mla{\relax}
\def\draft{
\def\thtystars{******************************}
\def\sixtystars{\thtystars\thtystars}
\typeout{}
\typeout{\sixtystars**}
\typeout{* Draft mode!
         For final version remove \protect\draft\space in source file *}
\typeout{\sixtystars**}
\typeout{}
\def\draftdate{\today}
\def\mua{\marginpar[\boldmath\hfil$\uparrow$]%
                   {\boldmath$\uparrow$\hfil}%
                    \typeout{marginpar: $\uparrow$}\ignorespaces}
\def\mda{\marginpar[\boldmath\hfil$\downarrow$]%
                   {\boldmath$\downarrow$\hfil}%
                    \typeout{marginpar: $\downarrow$}\ignorespaces}
\def\mla{\marginpar[\boldmath\hfil$\rightarrow$]%
                   {\boldmath$\leftarrow $\hfil}%
                    \typeout{marginpar: $\leftrightarrow$}\ignorespaces}
\def\Mua{\marginpar[\boldmath\hfil$\Uparrow$]%
                   {\boldmath$\Uparrow$\hfil}%
                    \typeout{marginpar: $\Uparrow$}\ignorespaces}
\def\Mda{\marginpar[\boldmath\hfil$\Downarrow$]%
                   {\boldmath$\Downarrow$\hfil}%
                    \typeout{marginpar: $\Downarrow$}\ignorespaces}
\def\Mla{\marginpar[\boldmath\hfil$\Rightarrow$]%
                   {\boldmath$\Leftarrow $\hfil}%
                    \typeout{marginpar: $\Leftrightarrow$}\ignorespaces}
\def\mpar##1{\marginpar{\hbadness10000%
                      \sloppy\hfuzz10pt\boldmath\bf##1}%
                      \typeout{marginpar: ##1}\ignorespaces}

\overfullrule 5pt
\oddsidemargin -15mm
\marginparwidth 29mm
}

\begin{document}

\thispagestyle{empty}
\def\thefootnote{\fnsymbol{footnote}}
\setcounter{footnote}{1}
\null
\draftdate
\hfill   TTP05-16\\
\strut\hfill  SFB/CPP-05-44\\
\strut\hfill hep-ph/0508253
\vskip 0cm
\vfill
\begin{center}
{\Large \bf
Electroweak corrections to hadronic photon production at large transverse momenta
\par}
\vskip 1em
{\large
{\sc Johann~H.~K\"uhn\footnote{Johann.Kuehn@physik.uni-karlsruhe.de}, 
A.~Kulesza\footnote{ania@particle.uni-karlsruhe.de},
S.~Pozzorini\footnote{pozzorin@particle.uni-karlsruhe.de},
M.~Schulze\footnote{schulze@particle.uni-karlsruhe.de} }}
\\[.5cm]
{\it Institut f\"ur Theoretische Teilchenphysik, 
Universit\"at Karlsruhe \\
D-76128 Karlsruhe, Germany}
\par
\end{center}\par
\vskip 1.0cm 
\vfill 
{\bf Abstract:} \par 

We study the impact of electroweak radiative corrections on 
direct production of photons with high transverse momenta  
at hadron colliders.
Analytic results for the weak one-loop corrections 
to the parton scattering reaction $\bar q q \to \gamma g$
and its crossed variants are presented.
For the high-energy region, where the corrections are strongly enhanced 
by logarithms of $\shat/M_W^2$, we derive simple asymptotic expressions
which approximate the exact one-loop results with high precision.
The dominant two-loop electroweak contributions are also calculated.
Numerical results are presented for the LHC and the Tevatron.
The corrections are negative and their size increases with transverse momentum.
For the LHC, where transverse momenta of 2 TeV or more can be reached,
the size of the one- and two-loop effects amounts up to
$-17\%$ and +3\%, respectively.
At the Tevatron, with transverse momenta up to 400 GeV, 
the one-loop corrections do not exceed $-4\%$ and the two-loop effects are negligible.
Finally we compare the cross sections for hadronic production of 
photons and $Z$ bosons and find that the electroweak corrections have 
an important impact on their ratio.


\par
\vskip 1cm
\noindent
August 2005 
\par
\null
\setcounter{page}{0}
\clearpage
\def\thefootnote{\arabic{footnote}}
\setcounter{footnote}{0}

\newcommand{\diagtI}[1]{
\begin{picture}(94.,90.)(-47,-45)
\ArrowLine(0,20)(-40,20)
\Vertex(0,20){1.5}
\Photon(0,20)(40,20){2.5}{6}
\ArrowLine(-40,-20)(0,-20)
\Vertex(0,-20){1.5}
\Gluon(0,-20)(40,-20){-2.5}{6}
\ArrowLine(0,-20)(0,20)
\Text(0,-35)[t]{#1}
\end{picture}}

\newcommand{\diagtII}[1]{
\begin{picture}(94.,90.)(-47,-45)
\ArrowLine(0,20)(-40,20)
\Vertex(0,20){1.5}
\Gluon(0,20)(40,20){2.5}{6}
\ArrowLine(-40,-20)(0,-20)
\Vertex(0,-20){1.5}
\Photon(0,-20)(40,-20){-2.5}{6}
\ArrowLine(0,-20)(0,20)
\Text(0,-35)[t]{#1}
\end{picture}}

\newcommand{\diagcI}[1]{
\begin{picture}(94.,90.)(-47,-45)
\ArrowLine(0,20)(-40,20)
\Photon(0,20)(40,20){2.5}{6}
\ArrowLine(-40,-20)(0,-20)
\Vertex(0,-20){1.5}
\Gluon(0,-20)(40,-20){-2.5}{6}
\ArrowLine(0,-20)(0,0)\ArrowLine(0,0)(0,20)
\Text(0,-35)[t]{#1}
\BCirc(0,0){4}\Line(-2.83,-2.83)(2.83,2.83)\Line(-2.83,2.83)(2.83,-2.83)
\end{picture}}

\newcommand{\diagcII}[1]{
\begin{picture}(94.,90.)(-47,-45)
\ArrowLine(0,20)(-40,20)
\Vertex(0,20){1.5}
\Gluon(0,20)(40,20){2.5}{6}
\ArrowLine(-40,-20)(0,-20)
\Vertex(0,-20){1.5}
\Photon(0,-20)(40,-20){-2.5}{6}
\ArrowLine(0,-20)(0,0)\ArrowLine(0,0)(0,20)
\Text(0,-35)[t]{#1}
\BCirc(0,0){4}\Line(-2.83,-2.83)(2.83,2.83)\Line(-2.83,2.83)(2.83,-2.83)
\end{picture}}

\newcommand{\diagcIII}[1]{
\begin{picture}(94.,90.)(-47,-45)
\ArrowLine(0,20)(-40,20)
\Photon(0,20)(40,20){2.5}{6}
\ArrowLine(-40,-20)(0,-20)
\Vertex(0,-20){1.5}
\Gluon(0,-20)(40,-20){-2.5}{6}
\ArrowLine(0,-20)(0,20)
\Text(0,-35)[t]{#1}
\BCirc(0,-20){4}\Line(-2.83,-22.83)(2.83,-17.18)\Line(-2.83,-17.18)(2.83,-22.83)
\end{picture}}

\newcommand{\diagcIV}[1]{
\begin{picture}(94.,90.)(-47,-45)
\ArrowLine(0,20)(-40,20)
\Vertex(0,20){1.5}
\Gluon(0,20)(40,20){2.5}{6}
\ArrowLine(-40,-20)(0,-20)
\Vertex(0,-20){1.5}
\Photon(0,-20)(40,-20){-2.5}{6}
\ArrowLine(0,-20)(0,20)
\Text(0,-35)[t]{#1}
\BCirc(0,20){4}\Line(-2.82,17.18)(2.83,22.82)\Line(-2.83,22.83)(2.83,17.18)
\end{picture}}

\newcommand{\diagcV}[1]{
\begin{picture}(94.,90.)(-47,-45)
\ArrowLine(0,20)(-40,20)
\Photon(0,20)(40,20){2.5}{6}
\ArrowLine(-40,-20)(0,-20)
\Vertex(0,-20){1.5}
\Gluon(0,-20)(40,-20){-2.5}{6}
\ArrowLine(0,-20)(0,20)
\Text(0,-35)[t]{#1}
\BCirc(0,20){4}\Line(-2.82,17.18)(2.83,22.82)\Line(-2.83,22.83)(2.83,17.18)
\end{picture}}

\newcommand{\diagcVI}[1]{
\begin{picture}(94.,90.)(-47,-45)
\ArrowLine(0,20)(-40,20)
\Vertex(0,20){1.5}
\Gluon(0,20)(40,20){2.5}{6}
\ArrowLine(-40,-20)(0,-20)
\Vertex(0,-20){1.5}
\Photon(0,-20)(40,-20){-2.5}{6}
\ArrowLine(0,-20)(0,20)
\Text(0,-35)[t]{#1}
\BCirc(0,-20){4}\Line(-2.83,-22.83)(2.83,-17.18)\Line(-2.83,-17.18)(2.83,-22.83)
\end{picture}}

\newcommand{\diagsI}[1]{
\begin{picture}(94.,90.)(-47,-45)
\ArrowLine(0,20)(-40,20)
\Vertex(0,20){1.5}
\Photon(0,20)(40,20){2.5}{6}
\ArrowLine(-40,-20)(0,-20)
\Vertex(0,-20){1.5}
\Gluon(0,-20)(40,-20){-2.5}{6}
\PhotonArc(0,0)(10,90,270){-2.5}{4}\Text(-15,0)[r]{$\scriptstyle{V}$}
\ArrowLine(0,-20)(0,-10)\Vertex(0,-10){1.5}\ArrowLine(0,-10)(0,10)\Vertex(0,10){1.5}\ArrowLine(0,10)(0,20)
\Text(0,-35)[t]{#1}
\end{picture}}

\newcommand{\diagsII}[1]{
\begin{picture}(94.,90.)(-47,-45)
\ArrowLine(0,20)(-40,20)
\Vertex(0,20){1.5}
\Gluon(0,20)(40,20){2.5}{6}
\ArrowLine(-40,-20)(0,-20)
\Vertex(0,-20){1.5}
\Photon(0,-20)(40,-20){-2.5}{6}
\PhotonArc(0,0)(10,90,270){-2.5}{4}\Text(-15,0)[r]{$\scriptstyle{V}$}
\ArrowLine(0,-20)(0,-10)\Vertex(0,-10){1.5}\ArrowLine(0,-10)(0,10)\Vertex(0,10){1.5}\ArrowLine(0,10)(0,20)
\Text(0,-35)[t]{#1}
\end{picture}}

\newcommand{\diagvI}[1]{
\begin{picture}(94.,90.)(-47,-45)
\ArrowLine(0,20)(-40,20)
\Vertex(0,20){1.5}
\Photon(0,20)(40,20){2.5}{6}
\ArrowLine(-40,-20)(-15,-20)
\Vertex(-15,-20){1.5}
\Vertex(15,-20){1.5}
\Gluon(15,-20)(40,-20){-2.5}{3.5}
\ArrowLine(-15,-20)(15,-20)
\Photon(-15,-20)(0,5){2.5}{4.5}\Text(-13,-5)[r]{$\scriptstyle{V}$}\ArrowLine(15,-20)(0,5)
\Vertex(0,5){1.5}\ArrowLine(0,5)(0,20)
\Text(0,-35)[t]{#1}
\end{picture}}

\newcommand{\diagvII}[1]{
\begin{picture}(94.,90.)(-47,-45)
\ArrowLine(-15,20)(-40,20)
\Vertex(-15,20){1.5}
\Vertex(15,20){1.5}
\Gluon(15,20)(40,20){2.5}{3.5}
\ArrowLine(15,20)(-15,20)
\Photon(0,-5)(-15,20){2.5}{4.5}\Text(-13,5)[r]{$\scriptstyle{V}$}\ArrowLine(0,-5)(15,20)
\Vertex(0,-5){1.5}\ArrowLine(0,-20)(0,-5)
\ArrowLine(-40,-20)(0,-20)
\Vertex(0,-20){1.5}
\Photon(0,-20)(40,-20){-2.5}{6}
\Text(0,-35)[t]{#1}
\end{picture}}

\newcommand{\diagvIII}[1]{
\begin{picture}(94.,90.)(-47,-45)
\ArrowLine(-15,20)(-40,20)
\Vertex(-15,20){1.5}
\Vertex(15,20){1.5}
\Photon(15,20)(40,20){2.5}{3.5}
\ArrowLine(15,20)(-15,20)
\Photon(0,-5)(-15,20){2.5}{4.5}\Text(-13,5)[r]{$\scriptstyle{V}$}\ArrowLine(0,-5)(15,20)
\Vertex(0,-5){1.5}\ArrowLine(0,-20)(0,-5)
\ArrowLine(-40,-20)(0,-20)
\Vertex(0,-20){1.5}
\Gluon(0,-20)(40,-20){-2.5}{6}
\Text(0,-35)[t]{#1}
\end{picture}}

\newcommand{\diagvIV}[1]{
\begin{picture}(94.,90.)(-47,-45)
\ArrowLine(0,20)(-40,20)
\Vertex(0,20){1.5}
\Gluon(0,20)(40,20){2.5}{6}
\ArrowLine(-40,-20)(-15,-20)
\Vertex(-15,-20){1.5}
\Vertex(15,-20){1.5}
\Photon(15,-20)(40,-20){-2.5}{3.5}
\ArrowLine(-15,-20)(15,-20)
\Photon(-15,-20)(0,5){2.5}{4.5}\Text(-13,-5)[r]{$\scriptstyle{V}$}\ArrowLine(15,-20)(0,5)
\Vertex(0,5){1.5}\ArrowLine(0,5)(0,20)
\Text(0,-35)[t]{#1}
\end{picture}}

\newcommand{\diagvV}[1]{
\begin{picture}(94.,90.)(-47,-45)
\ArrowLine(-15,20)(-40,20)
\Vertex(-15,20){1.5}
\Vertex(15,20){1.5}
\Photon(15,20)(40,20){2.5}{3.5}
\Photon(15,20)(-15,20){-2.5}{4.5}\Text(0,25)[b]{$\scriptstyle{W^\pm}$}
\ArrowLine(0,-5)(-15,20)\Photon(0,-5)(15,20){2.5}{4.5}\Text(13,5)[l]{$\scriptstyle{W^\mp}$}
\Vertex(0,-5){1.5}\ArrowLine(0,-20)(0,-5)
\ArrowLine(-40,-20)(0,-20)
\Vertex(0,-20){1.5}
\Gluon(0,-20)(40,-20){-2.5}{6}
\Text(0,-35)[t]{#1}
\end{picture}}

\newcommand{\diagvVI}[1]{
\begin{picture}(94.,90.)(-47,-45)
\ArrowLine(0,20)(-40,20)
\Vertex(0,20){1.5}
\Gluon(0,20)(40,20){2.5}{6}
\ArrowLine(-40,-20)(-15,-20)
\Vertex(-15,-20){1.5}
\Vertex(15,-20){1.5}
\Photon(15,-20)(40,-20){-2.5}{3.5}
\Photon(-15,-20)(15,-20){-2.5}{4.5}\Text(0,-25)[t]{$\scriptstyle{W^\pm}$}
\ArrowLine(-15,-20)(0,5)\Photon(15,-20)(0,5){2.5}{4.5}\Text(13,-5)[l]{$\scriptstyle{W^\mp}$}
\Vertex(0,5){1.5}\ArrowLine(0,5)(0,20)
\Text(0,-35)[t]{#1}
\end{picture}}

\newcommand{\diagbI}[1]{
\begin{picture}(94.,90.)(-47,-45)
\ArrowLine(-15,20)(-40,20)
\Vertex(-15,20){1.5}
\Vertex(15,20){1.5}
\Photon(15,20)(40,20){2.5}{3.5}
\ArrowLine(15,20)(-15,20)
\ArrowLine(-40,-20)(-15,-20)
\Vertex(-15,-20){1.5}
\Vertex(15,-20){1.5}
\Gluon(15,-20)(40,-20){-2.5}{3.5}
\ArrowLine(-15,-20)(15,-20)
\Photon(-15,-20)(-15,20){2.5}{5}\Text(-20,0)[r]{$\scriptstyle{V}$}
\ArrowLine(15,-20)(15,20)
\Text(0,-35)[t]{#1}
\end{picture}}

\newcommand{\diagbII}[1]{
\begin{picture}(94.,90.)(-47,-45)
\ArrowLine(-15,20)(-40,20)
\Vertex(-15,20){1.5}
\Vertex(15,20){1.5}
\Gluon(15,20)(40,20){2.5}{3.5}
\ArrowLine(15,20)(-15,20)
\ArrowLine(-40,-20)(-15,-20)
\Vertex(-15,-20){1.5}
\Vertex(15,-20){1.5}
\Photon(15,-20)(40,-20){-2.5}{3.5}
\ArrowLine(-15,-20)(15,-20)
\Photon(-15,-20)(-15,20){2.5}{5}\Text(-20,0)[r]{$\scriptstyle{V}$}
\ArrowLine(15,-20)(15,20)
\Text(0,-35)[t]{#1}
\end{picture}}

\newcommand{\diagbIII}[1]{
\begin{picture}(94.,90.)(-47,-45)
\ArrowLine(-15,20)(-40,20)
\Vertex(-15,20){1.5}
\Vertex(15,20){1.5}
\Photon(15,20)(40,20){2.5}{3.5}
\Photon(15,20)(-15,20){-2.5}{4.5}\Text(0,25)[b]{$\scriptstyle{W^\pm}$}
\ArrowLine(-40,-20)(-15,-20)
\Vertex(-15,-20){1.5}
\Vertex(15,-20){1.5}
\Gluon(15,-20)(40,-20){-2.5}{3.5}
\ArrowLine(-15,-20)(15,-20)
\Photon(-15,-20)(15,20){2.5}{7}\Text(13,0)[lb]{$\scriptstyle{W^\mp}$}
\ArrowLine(15,-20)(-15,20)
\Text(0,-35)[t]{#1}
\end{picture}}

\newcommand{\diagqqzct}{
\begin{picture}(40.,20.)(0,-3)
\ArrowLine(0,0)(-20,20)
\Photon(0,0)(30,0){2.5}{4}
\ArrowLine(-20,-20)(0,0)
\BCirc(0,0){4}\Line(-2.83,-2.83)(2.83,2.83)\Line(-2.83,2.83)(2.83,-2.83)
\end{picture}}

\newcommand{\diagqqgct}{
\begin{picture}(40.,20.)(0,-3)
\ArrowLine(0,0)(-20,20)
\Gluon(0,0)(30,0){2.5}{4}
\ArrowLine(-20,-20)(0,0)
\BCirc(0,0){4}\Line(-2.83,-2.83)(2.83,2.83)\Line(-2.83,2.83)(2.83,-2.83)
\end{picture}}

\newcommand{\diagqqct}{
\begin{picture}(40.,10.)(0,-3)
\ArrowLine(0,0)(-30,0)
\ArrowLine(30,0)(0,0)
\BCirc(0,0){4}\Line(-2.83,-2.83)(2.83,2.83)\Line(-2.83,2.83)(2.83,-2.83)
\end{picture}}

\newpage

\section{Introduction}
The study of direct photon production, consisting of the QCD Compton
process $gq \rightarrow \gamma q$ and the annihilation process $\bar q
q \rightarrow \gamma g$, has always been an important topic in
hadron colliders physics, both theoretically~\cite{Fritzsch:1977eq,Aurenche:1983ws}
and experimentally~\cite{Abe:1994rr}.
Since the photons do not fragment and can be clearly identified
experimentally, direct photon production provides a 
much clearer probe of the hard-scattering dynamics than jet production
processes. Therefore the study of large transverse momentum direct (prompt)
photon production constitutes an important test 
of perturbative QCD and the point-like nature of quarks and gluons.
Furthermore, the reaction contributes to background for 
many signals of new physics.
Being embedded in the
environment of hadronic collisions, the reaction necessarily
involves hadronic physics, like parton distributions, and
depends on the strong coupling constant. 
In turn, the cross
section for direct production of photons, their tranverse momentum ($p_\rT$)
and rapidity distributions can be used to gauge the parton distribution
functions. Since the gluon distribution enters already at the leading
order, the measurement of the direct photon production is an important means 
to constrain information on the gluon content of the
proton~\cite{Aurenche:1989gv}.
In particular, large transverse momentum production provides a unique
opportunity for determination of gluon densities at large $x$.

Apart from the direct process, prompt photons 
can be also produced through a fragmentation process. However, 
most of the fragmentation contribution can be removed by applying an
isolation criterion. The importance of the remaining contribution from 
fragmentation, after applying the isolation cut, is expected to
decrease with higher $p_\rT$. Moreover, background
processes to isolated direct photon production, i.e. photon production 
through decays of neutral mesons ($\pi^0$, $\eta$) coming from jet
fragmentation, are shown to be less important at large $p_\rT$~\cite{Kumar:2003ue}.
To achieve reliable predictions at high $p_\rT$, QCD corrections in
next-to-leading order~\cite{Aurenche:1983ws}, are mandatory.
The corrections can amount to several tens of percent~\cite{Aurenche:1983ws,Kumar:2003ue,Huston:1995vb}, depending on
the observable under consideration, value of $p_\rT$, and details of the
calculation such as jet definition or 
renormalization and factorization scales.
The evaluation of next-to-next-to-leading order
corrections involves two-loop virtual plus a variety of combined
virtual plus real corrections and is a topic presently pursued by
various groups~\cite{Gehrmann-DeRidder:2004xe}.

For the experiments at the Large Hadron Collider (LHC) a new
aspect comes into play. The high center-of-mass energy in
combination
with the enormous luminosity will allow to explore
parton-parton
scattering up to 
energies of several TeV 
and correspondingly production of gauge
bosons with transverse momenta up to \mbox{2 TeV} or even beyond. 
In this region electroweak
corrections from virtual weak boson exchange increase
strongly, with the dominant terms 
in $L$-loop approximation being 
leading logarithms of the form 
$\alpha^L\log^{2L}(\hat{s}/M_W^2)$,
next-to-leading logarithms of the form 
$\alpha^L\log^{2L-1}(\hat{s}/M_W^2)$, and so on.
These corrections, also known as electroweak Sudakov logarithms, may
well amount to several tens of percent.
They have been studied in
great detail for processes involving fermions in
\citeres{Kuhn:1999de,Pozzorini:2004rm}. 
Investigations on the
dominant and the next-to-leading logarithmic terms are also
available for reactions involving gauge and Higgs bosons
\cite{Denner:2001jv,Pozzorini:rs,Denner:2003wi}.
A recent survey of the literature on 
logarithmic electroweak corrections
can be found in \citere{Hollik:2004dz}.
The impact of these corrections
at hadron colliders has been studied in \citere{Accomando:2001fn}.
Specifically, hadronic $Z$-boson production at large $p_\rT$ has
been investigated in next-to-leading logarithmic 
approximation, including the two-loop terms \cite{Kuhn:2004em,Kuhn:2005az}. 
Numerical results for the complete one-loop terms 
have been presented in \citere{Maina:2004rb} both for $Z$ and photon 
production.

It is the aim of this work to obtain an independent evaluation of
the complete one-loop weak 
corrections to the photon production at large transverse momentum, 
and to present the full result in analytic form. At
the partonic level
the reactions $\bar{q} q\rightarrow\gamma g$, 
$g q\rightarrow\gamma q$ and
$g \bar{q}\rightarrow\gamma \bar{q}$ 
with $q=u,d,c,s$ or $b$ have to
be considered which are, however, trivially related by crossing
and appropriate exchange of coupling constants. 
Much of the calculation proceeds in close similarity with the
evaluation of $Z$ production presented in \citere{Kuhn:2005az}.
In fact, a significant part of the result can be obtained directly 
from the expressions published in~\citere{Kuhn:2005az}.
We split the
corrections into an ''Abelian'' and a ''non-Abelian'' 
component. 
We present analytic results for the exact one-loop corrections
that permit to predict
separately the various quark-helicity contributions.
We also derive compact analytic expressions for the high-energy behaviour 
of the  corrections.
Here we include quadratic and linear logarithms as well as those 
terms that are not logarithmically enhanced at high energies but 
neglect all contributions of $\ord(M^2_W/\hat{s})$. 
The accuracy of this approximation is then discussed.

After convolution with parton distribution functions,
radiatively corrected predictions for transverse momentum distributions
of photons at hadron colliders are obtained.
Concerning perturbative QCD, these predictions are based on the lowest
order and thus proportional to $\alpha_\rS$.
To obtain realistic cross sections,
higher-order QCD corrections would have to be included, in 
next-to-leading or even next-to-next-to-leading order.

The paper is organized as follows: In \refse{se:kinematics} the Born
approximation, our conventions and the kinematics are
introduced. \refse{se:corr} contains
a detailed description of the radiative corrections. 
\refse{se:renormalisation} is concerned with the renormalization procedure. 
In \refse{se:results} our one-loop results are given in terms of
expressions that had been introduced in \citere{Kuhn:2005az} for the
case of $Z$-boson production.
Special attention is paid to results in the high-energy limit
(\refse{se:helimit}),
which can be cast into a compact form.
\refse{se:twoloops} contains results for the dominant two-loop terms,
i.e. the leading and next-to-leading Sudakov logarithms.
In \refse{se:numerics} a detailed discussion of numerical results is presented,
both for $pp$ and $p\bar p$ collisions at 14 TeV and 2 TeV, respectively.
The leading and next-to-leading two-loop logarithmic terms 
are included in this numerical analysis.
\refse{se:conc} contains a brief summary.

\section{Preliminaries}
\label{se:kinematics}
The  transverse momentum ($\pT$) distribution of photons in the reaction 
$h_1 h_2 \to \gamma + \mathrm{jet}$ 
is given by 
\newcommand{\pdf}[4]{f_{#1,#2}(#3,#4)}
\beq
\label{hadroniccs}
\frac{\rd \si^{h_1 h_2}}{\rd \pT}=
\sum_{i,j}\int_0^1\rd x_1 \int_0^1\rd x_2
\;\theta(x_1 x_2-\hat\tau_{\rm min})
\pdf{h_1}{i}{x_1}{\mu^2}
\pdf{h_2}{j}{x_2}{\mu^2}
\frac{\rd \hat{\si}^{i j}}{\rd \pT}
,  
\eeq
where 
$\hat \tau_{\rm min} =4 \pT^2/s$
and $\sqrt{s}$ is the collider energy.
The indices  $i, j$  denote initial state partons ($q, \bar q, g$) and
$f_{h_1,i}(x,\mu^2)$, $f_{h_2,j}(x,\mu^2)$
are the corresponding parton distribution functions.
The partonic cross section for 
the subprocess $i j \to \gamma k$ is denoted by $\hat {\si}^{ij}$ and
the sum in~(\ref{hadroniccs}) 
runs over all $i,j$ combinations corresponding to the subprocesses
\beq\label{processes}
\bar q q \to \gamma g,\quad
q\bar q\to \gamma g,\quad
g q \to \gamma  q,\quad
q g \to \gamma  q,\quad
\bar q g  \to \gamma  \bar q,\quad
g \bar q   \to \gamma  \bar q
\eeq
with $q=u,d,c,s$ or $b$.
The Mandelstam variables for the subprocess $i j \to \gamma k$  are defined
in the standard way
\beq
\shat=(p_i+ p_j)^2 
,\qquad 
\that=(p_i- p_\gamma)^2 
,\qquad 
\uhat=(p_j-p_\gamma)^2
. 
\eeq
The momenta  $p_{i}$, $p_{j}$, $p_{k}$ and $p_\gamma$ of the partons are massless.
In terms of $x_1,x_2,\pT$ and the collider energy $\sqrt{s}$ we have
\beq
\shat=x_1 x_2 s,\qquad
\that=- \frac{\shat}{2}(1-\cos\theta),\qquad
\uhat=- \frac{\shat}{2}(1+\cos\theta),
\eeq
with $\cos\theta=\sqrt{1- 4\pT^2 /\shat}$
corresponding to the cosine of the angle between the momenta $p_i$ and
$p_\gamma$ in the partonic center-of-mass frame.

The angular and the $\pT$ distribution 
for the unpolarized partonic subprocess $ij\to \gamma k$
read 
\beqar\label{partoniccs0}
\frac{\rd \hat{\si}^{i j}}{\rd \cos\theta}
&=&
\frac{1}{32\pi N_{i j}\shat}
\;\overline{\sum}|\M^{i j}|^2
\eeqar
and
\beqar\label{partoniccs1}
\frac{\rd \hat{\si}^{i j}}{\rd \pT}
&=&
\frac{\pT}{8\pi N_{i j}\shat|\that-\uhat|}
\left[
\overline{\sum}|\M^{i j}|^2+(\that\leftrightarrow \uhat)
\right]
,
\eeqar
where
$
\overline{\sum}=
\frac{1}{4}
\sum_{\mathrm{pol}} 
\sum_{\mathrm{col}} 
$
involves the sum over polarization and color as well as the average factor $1/4$ for initial-state polarization. 
The factor $1/N_{i j}$ in \refeq{partoniccs0}-\refeq{partoniccs1}, 
with 
$N_{\bar q q}=N_{q\bar q}=N_\rc^2$, 
$N_{g q}=N_{q g}=N_{\bar q g}=N_{g\bar q}=N_\rc (N_\rc^2-1)$, 
and $N_\rc=3$, accounts for  the initial-state colour average.

In the following we present analytic results for the 
unpolarized squared matrix elements $\overline{\sum} |\M^{\qbar q}|^2$
for the $\bar q q\to \gamma g$ process. Corresponding results for the 
other processes in \refeq{processes} are easily obtained by means of 
CP symmetry and crossing transformations \cite{Kuhn:2005az}.

{\unitlength 1pt \small
\begin{figure}
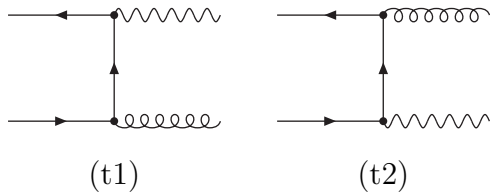

\begin{center}
\diagtI{(t1)}
\diagtII{(t2)}
\end{center}
 \caption{Tree-level Feynman diagrams for the process $\bar q q \to \gamma g$.}
 \label{fig:treediags}
 \end{figure}}

The lowest-order contribution results from the tree diagrams depicted in 
\reffi{fig:treediags} and reads
\beq\label{generalamplitude}
\overline{\sum}|\M_0^{\bar q q}|^2=
16 \pi^2 \alpha \alpha_\rS (N_\rc^2-1)
Q_{q}^2 \,
\frac{\that^2+\uhat^2}{\that\uhat} 
,
\eeq
where $Q_u=2/3$, $Q_d=-1/3$, $\alpha=e^2/(4 \pi)$ and $\alpha_\rS=g_\rS^2/(4 \pi)$ 
are the quark electric charges, and the electromagnetic and the strong
coupling constants, respectively.

\section{One-loop corrections}
\label{se:corr}

In this section, we present analytic results for the one-loop weak corrections to the  process $\qbar q \to \gamma g$.
They are closely related to those presented in 
\citere{Kuhn:2005az} for the  process $\qbar q \to Z g$. 
In fact, the one-loop diagrams for  $\qbar q \to \gamma g$, see
\reffi{fig:loopdiags}, are analogous to the diagrams for  $\qbar q \to Z g$.  
Hence a large subset of corrections 
to  $\qbar q \to \gamma g$ can be obtained directly from 
the results for $\qbar q \to Z g$ by setting $M_Z$ to zero.
However, the renormalization
counterterms need to be calculated separately for the  $\qbar q \to
\gamma g$ process.

Similarly as in the $Z$-boson production case \cite{Kuhn:2005az}, we do not include
electromagnetic corrections in the full one-loop result. The 
renormalized weak corrections are infrared finite. 
To obtain the results of \citere{Kuhn:2005az}, we assumed all
quarks to be massless, neglected diagrams involving couplings of
quarks to Higgs bosons or would-be-Goldstone bosons and
omitted quark-mixing effects. The calculation of \citere{Kuhn:2005az}
was performed at the level of matrix elements and allows for a full
control over polarization effects. The one-loop amplitude was split
into two parts according to
the structure of the gauge group generators in front of each term:
Abelian (characteristic for Abelian theories) and non-Abelian
(originating from the non-commutativity of weak interactions). 
Tensor loop integrals, appearing in
the expressions for one-loop corrections were reduced to scalar integrals
by means of the Passarino-Veltman technique.  Assuming
zero quark masses led to 
mass singularities of collinear nature, present at the level of
individual integrals.
However, these cancel between various scalar integrals, making it possible to write the final result in terms of
finite combinations of the integrals. Further details concerning
the calculation can be found in \citere{Kuhn:2005az}.

In the following, only these contributions  
are presented in detail that cannot be directly derived 
from  \citere{Kuhn:2005az} by setting $M_Z$ to zero.
We start discussing renormalization in the on-shell scheme, in \refse{se:renormalisation}.
The complete one-loop corrections 
and their asymptotic behaviour are then presented in \refse{se:results} and \refse{se:helimit}, respectively.

{\unitlength 1pt \small
\begin{figure}
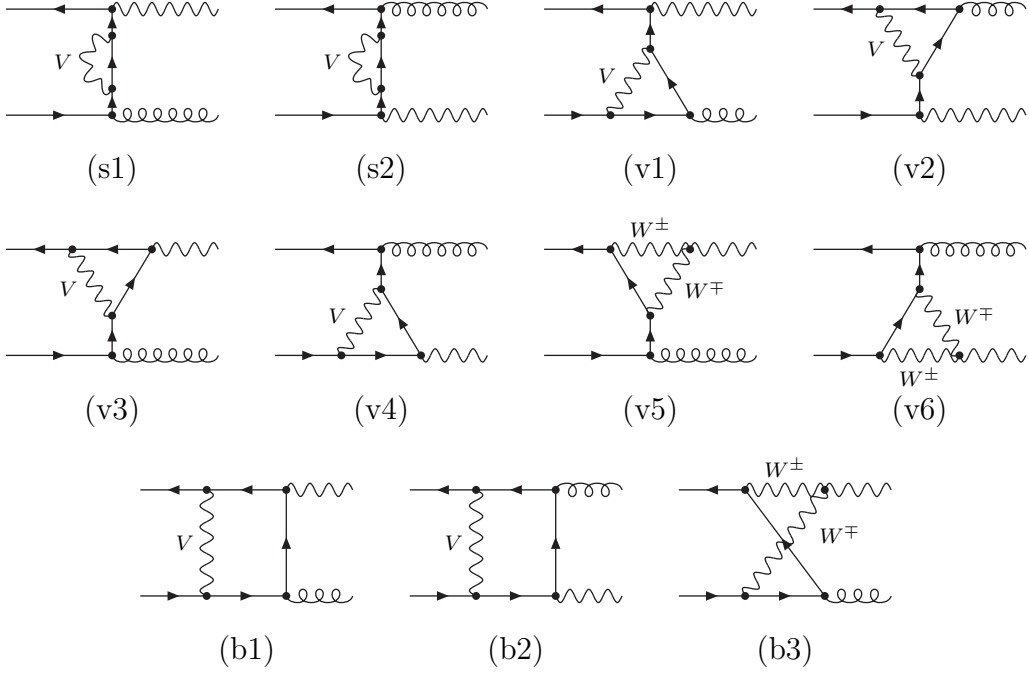

\begin{center}
\diagsI{(s1)}
\diagsII{(s2)}
\diagvI{(v1)}
\diagvII{(v2)}
\diagvIII{(v3)}
\diagvIV{(v4)}
\diagvV{(v5)}
\diagvVI{(v6)}
\diagbI{(b1)}
\diagbII{(b2)}
\diagbIII{(b3)}
\end{center}
 \caption{
One-loop Feynman diagrams for the process $\bar q q \to \gamma g$.
The diagrams  v5, v6 and b3 involve only 
charged weak bosons, $W^\pm$, whereas the other diagrams receive contributions from neutral and charged 
weak bosons, $V=Z,W^\pm$.
}
 \label{fig:loopdiags}
 \end{figure}}

\subsection{Renormalization}
\label{se:renormalisation}
\newcommand{\Deltamsbar}{\bar \Delta_{\mathrm{UV}}}
The renormalization of the $\qbar q\to \gamma g$ process is
provided by the $\gamma q\qbar$ counterterm.%
\footnote{
Same as for the case of the  $\qbar q\to Z g$ process \cite{Kuhn:2005az},
the counterterm contributions associated with the  $gq\qbar$ vertex
and the quark propagator cancel.
}
This can be written, similarly to the  $Z q\qbar$ counterterm \cite{Kuhn:2005az}, as
\beqar\label{qqzct}
\diagqqzct
&=& -\ri e \gamma^{\mu}
\sum_{\lambda=\mathrm{R,L}} \omega_{\lambda}  \left[Q_{q} \, 
\left(\delta Z_{q_{\lambda}}+\delta C^{\mathrm{A}}_{q_\lambda}\right)
+ T^3_{q_{\lambda}} \, \delta C^{\mathrm{N}}_{q_{\lambda}}
\right]
\eeqar
\vspace{5mm}
\pagebreak

\noindent
with 
$\omega_\rR=(1+\gamma^5)/2$ and
$\omega_\rL=(1-\gamma^5)/2$. 
As in  \citere{Kuhn:2005az}, 
\beq\label{Qwfcts}
\delta Z_{q_{\lambda}} =
\frac{\alpha}{4\pi} \sum_{V=\mathrm{Z,W^\pm}}
\left( I^V I^{\bar{V}} \right)_{q_{\lambda}}
\left[\frac{3}{2} -\frac{A_0(M_V^2)}{M_V^2}\right]
\eeq
are the wave-function renormalization constants 
of massless chiral quarks.
The coupling factors in \refeq{Qwfcts} read
\beqar
\label{couplfact1}
\left(I^Z I^{Z}\right)_{q_{\lambda}}
&=&
\left(\frac{\cw}{\sw} T^3_{q_\la}-\frac{\sw}{\cw}\frac{Y_{q_\la}}{2}
\right)^2
,\qquad
\sum_{V=W^{\pm}} \left(I^V I^{\bar{V}}\right)_{q_{\lambda}}
=
\frac{\de_{\la\rL}}{2\sw^2}
\eeqar
with the shorthands $\cw=\cos{\theta_\rw}$ and  $\sw=\sin{\theta_\rw}$
for the  weak mixing angle $\theta_\rw$.
 $Y_{q_\la}/2=Q_{q}-T^3_{q_\la}$ and
$T^3_{q_\la}$ is the weak isospin of chiral quarks, \ie 
$T^3_{u_\la}=\de_{\la\rL}/2$ and $T^3_{d_\la}=-\de_{\la\rL}/2$.
The remaining counterterm contributions in \refeq{qqzct}
are split in an Abelian part, $\de C^{\mathrm{A}}_{q_\la}$, 
proportional to the tree-level coupling $Q_q$
and a non-Abelian part, $\de C^{\mathrm{N}}_{q_\la}$, proportional to 
the  weak isospin $T^3_{q_\la}$.
In the following we adopt the on-shell renormalization scheme \cite{Denner:1991kt}, 
with $\cw^2=1-\sw^2={M_W^2}/{M_Z^2}$ 
and the electromagnetic coupling constant 
defined in the Thompson limit, 
\ie at the scale zero.
In this scheme, 
the Abelian counterterm  $\de C^{\mathrm{A}}_{q_\la}$ vanishes owing to 
Ward identities \cite{Denner:1991kt},
\beqar\label{deCCtsa}
\delta C^{\mathrm{A}}_{q_\lambda} 
&=&
\frac{1}{2} \left( \delta Z_{AA} +\frac{\sw}{\cw} \delta Z_{ZA} 
+ \frac{\delta e^2}{e^2} \right)=0.
\eeqar
The expressions for the counterterms $\delta Z_{AA}$, $\delta
Z_{ZA}$ and ${\delta e^2}/{e^2}$ can be found in \citere{Denner:1991kt}.
As well known, the on-shell counterterm  ${\delta e^2}/{e^2}$
contains large logarithms of light-fermion masses, which 
are responsible for the running of $\alpha$ 
from the scale zero to the characteristic scale of the process. 
However, owing to the Ward identity,
for the case of on-shell photon production
these logarithms are cancelled by corresponding terms present 
in $\delta Z_{AA}$, see  \refeq{deCCtsa}.
This justifies our choice of $\alpha$ at the scale zero 
as input parameter.

The non-Abelian counterterm  reads 
\beqar\label{deCCtsb}
\delta C^{\mathrm{N}}_{q_{\lambda}} 
&=&
-\frac{1}{2\sw\cw}
\delta Z_{ZA},
\eeqar
where
\beq\label{Zwfcts}
\delta Z_{ZA} 
= 
2 \frac{\Sigma^{\mathrm{AZ}}_{\mathrm{T}}(0)}{M_Z^2}
=
\frac{\alpha}{4\pi}\frac{4 \cw}{\sw}\left[
\Deltamsbar-\log\left(\frac{M_W^2}{M_Z^2}\right)
\right]
\eeq
with 
\beq\label{msbarsubt}
\Deltamsbar = 1/\varepsilon -\gamma_{\mathrm{E}} +\log(4\pi)+\log\left(\frac{\mu_\rR^2}{M_Z^2}\right)
\eeq
in $D=4-2\varepsilon$ dimensions.
Here $\mu_\rR$ denotes the scale of dimensional
regularization. 

\subsection{Result}
\label{se:results}
The complete  $\ord(\alpha^2\alpha_\rS)$ result
for the $\qbar q\to \gamma g$ process can be written as
\beqar\label{generalresult}
\lefteqn{
\overline{\sum}
|\M^{\qbar q}_{1}|^2 =
8 \pi^2 \alpha \alpha_\rS (N_\mathrm{c}^2-1)\,
}\quad
\nl&&{}\times      
\sum_{\lambda=\mathrm{R,L}}  
\Bigg\{  
Q_{q}^2 \bigg[
H_0 + \frac{\alpha}{2\pi}
\sum_{V=\mathrm{Z,W^{\pm}}} \left(I^V I^{\bar{V}}\right)_{q_{\lambda}}
\,H_1^{\mathrm{A}}(M_V^2)
\bigg]
\nl&&{}      
+ T^3_{q_\la}Q_{q} 
\bigg[
2 H_0\delta C^{\mathrm{N}}_{q_{\lambda}}
+ \frac{\alpha}{2\pi}
\frac{1}{\sw^2}H_1^{\mathrm{N}}(M_W^2)
\bigg]\Bigg\}
,
\eeqar
where $H_0={(\that^2+\uhat^2)}/{(\that\uhat)}$ is the Born contribution and $\delta C^{\mathrm{N}}_{q_{\lambda}}$
is the counterterm specified in \refse{se:renormalisation}.
The remaining functions $H_1^{\mathrm{A/N}}(M_V^2)$
represent the Abelian (A) and non-Abelian (N) contributions 
resulting from the loop diagrams (see \reffi{fig:loopdiags}) 
involving virtual weak bosons, $V=Z,W^\pm$,
and the fermionic wave-function renormalization constants \refeq{Qwfcts}.
The above  result is analogous to the one presented in \citere{Kuhn:2005az}  for the $\qbar q\to Z g$ process. In particular, the functions 
$H_1^{\mathrm{A/N}}(M_V^2)$  for the  $\qbar q \to \gamma g$ process
can be obtained from the corresponding 
functions  for the $\qbar q\to Z g$ process
\cite{Kuhn:2005az} 
by setting the mass of the external $Z$ boson equal to zero.%
\footnote{
The functions  $H_1^{\mathrm{A/N}}(M_V^2)$
presented in  \citere{Kuhn:2005az} consist of linear combinations of
loop integrals $J_0,\dots,J_{14}$. 
We observe that the integrals
 $J_2,J_{8}$ and $J_{10}$ are logarithmically singular for $M_Z\to 0$.
However the corresponding coefficients are proportional 
to $M_Z^2$. Thus the contributions of such loop integrals vanish for 
$M_Z\to 0$.}
This simple relation between the loop corrections to the 
processes  $\qbar q \to \gamma g$ and $\qbar q \to Z g$ 
is due to the fact that, for $Z$ production,
the longitudinal contributions to the $Z$-boson polarization sum vanish.
This is a consequence of the Ward identities presented in 
Sect.~3.6 of \citere{Kuhn:2005az}.

\subsection{High-energy limit}
\label{se:helimit}

Let us now consider the high-energy region, $\shat/M_W^2\gg 1$, where 
the weak corrections are strongly enhanced by logarithms of the form
$\log(\shat/M_W^2)$.
As in  \citere{Kuhn:2005az} we derive 
compact analytic expressions that describe the 
asymptotic high-energy behaviour of the one-loop corrections, 
$H_1^{\mathrm{A/N}}(M_V^2)$, to 
next-to-next-to-leading logarithmic (NNLL) 
accuracy. This approximation includes 
double and single logarithms as well as 
those contributions that are not logarithmically enhanced
at high energies.
The NNLL expressions for  
$H_1^{\mathrm{N}}(M_W^2)$, 
$H_1^{\mathrm{A}}(M_W^2)$ and
$H_1^{\mathrm{A}}(M_Z^2)$,
obtained by means of 
general results for the high-energy limit of one-loop integrals \cite{Roth:1996pd}, have the form
\beq\label{nnllstracture}
H_1^{\mathrm{A/N}}(M_V^2)\NNLLa
\mathrm{Re }\,\left[
g_0^{\mathrm{A/N}}(M_V^2)\,
\frac{ \that^2+\uhat^2}{\that\uhat}
+g_1^{\mathrm{A/N}}(M_V^2)\,
\frac{ \that^2-\uhat^2}{\that\uhat}
+g_2^{\mathrm{A/N}}(M_V^2)
\right].
\eeq
In principle, the functions $g_i^{\mathrm{A}/\mathrm{N}}(M_V^2)$ 
for the   $\qbar q \to \gamma g$ process can be obtained 
from the corresponding functions for the   $\qbar q \to Z g$ process
\cite{Kuhn:2005az} by setting the mass of the external $Z$  boson equal to zero. However,  this must be done with care since the $Z$-mass dependence of the 
$g_i^{\mathrm{A}/\mathrm{N}}(M_V^2)$ functions in  \citere{Kuhn:2005az} is not shown explicitly.
For the case of photon production we find that 
the functions $g_i^{\mathrm{A}}(M_V^2)$   with $i=0,1,2$ and 
$g_j^{\mathrm{N}}(M_W^2)$ with $j=1,2$ are the same as
in  \citere{Kuhn:2005az}.
Instead, for the non-Abelian function 
$g_0^{\mathrm{N}}(M_W^2)$ we obtain 
\beqar\label{heres1}
 g_0^{\mathrm{N}}(M_W^2)&=&
2\left[\Deltamsbar-\log\left(\frac{M_W^2}{M_Z^2}\right)\right]
+\log^2\left(\frac{-\shat}{M_W^2}\right)
-\log^2\left(\frac{-\that}{M_W^2}\right)
-\log^2\left(\frac{-\uhat}{M_W^2}\right)
\nl&&{}
+\log^2\left(\frac{\that}{\uhat}\right)
-\frac{3}{2}\Biggl[
\log^2\left(\frac{\that}{\shat}\right)
+\log^2\left(\frac{\uhat}{\shat}\right)
\Biggr]
-2 \pi^2.
\eeqar
Here  the logarithmic terms are the same as in the $Z$-production case
whereas the constants are different. In particular, the coefficient of the 
$\pi^2$ term is different and no term of the form $\pi/\sqrt{3}$ is present 
in \refeq{heres1}.
Logarithms with negative arguments 
in \refeq{heres1} are defined through the 
usual $\ri\varepsilon$ prescription, 
$\rhat\to \rhat +\ri\varepsilon$
for  $\rhat=\shat,\that,\uhat$.
Assuming, as in \citere{Kuhn:2005az}, that one of the Mandelstam invariants 
is positive and the other two are negative, for the real part of \refeq{heres1}
we obtain
\beqar\label{heres1b}
\mathrm{Re }\, g_0^{\mathrm{N}}(M_W^2)&=&
2\left[\Deltamsbar-\log\left(\frac{M_W^2}{M_Z^2}\right)\right]
+\log^2\left(\frac{|\shat|}{M_W^2}\right)
-\log^2\left(\frac{|\that|}{M_W^2}\right)
-\log^2\left(\frac{|\uhat|}{M_W^2}\right)
\nl&&{}
+\log^2\left(\frac{|\that|}{|\uhat|}\right)
-\frac{3}{2}\Biggl[
\log^2\left(\frac{|\that|}{|\shat|}\right)
+\log^2\left(\frac{|\uhat|}{|\shat|}\right)
\Biggr]
-\frac{\pi^2}{2}\theta(-\shat).
\eeqar
This result applies directly to the $\qbar q\to \gamma g$  
process, 
where $\shat>0,\that<0$ and $\uhat<0$. In this case $\theta(-\shat)=0$. 
The $\theta$ term becomes non-vanishing when \refeq{heres1b}
is translated, 
by means of permutations of the Mandelstam invariants
(crossing transformations),
to the processes in \refeq{processes} that involve gluons in the initial state.

\section{Two-loop corrections}
\label{se:twoloops}
\newcommand{\logar}[2]{\mathrm{L}^{#1}_{#2}}

In the TeV energy region, 
the logarithmic electroweak corrections are very large.
The one-loop contributions can amount to tens of percent
and also the two-loop logarithmic terms can reach the level of several 
percent, see \eg \citere{Kuhn:2005az}. 
Such higher-order contributions are thus mandatory 
for precise theoretical predictions.
In the following, we present the  
two-loop corrections to the process  $\qbar q\to \gamma g$
to next-to-leading logarithmic (NLL) accuracy.
For a discussion of the calculation we refer to 
\citere{Kuhn:2004em}, where the same class of corrections 
has been computed for the process  $\qbar q\to Z g$.

The unpolarized squared matrix element for  $\qbar q\to \gamma g$,
including NLL terms 
up to the two-loop level, has the general form 
\beq\label{generalamplitudetwo}
\overline{\sum}|\M_2^{\bar q q}|^2=
8 \pi^2
{\alpha\alpha_\rS}
(N_\rc^2-1)
\frac{\that^2+\uhat^2}{\that\uhat} 
\left[
A^{(0)}
+\left(\frac{\alpha}{2\pi}\right)A^{(1)}
+\left(\frac{\alpha}{2\pi}\right)^2 A^{(2)}
\right]
.
\eeq
The Born contribution reads 
\beq\label{bornresult}
 A^{(0)}= 2 Q_{q}^2.
\eeq
At one loop, the NLL part consists of double- and single-logarithmic
terms and reads
\beq\label{oneloopresult}
A^{(1)}\NLLa - \sum_{\la=\rL,\rR}
Q_{q}\left[
Q_{q}\left(\cew_{q_\la}-Q_q^2\right)\left(\logar{2}{\shat}-3\logar{}{\shat}\right)
+
\frac{1}{\sw^2}T^3_{q_\la}
\left(\logar{2}{\that}+\logar{2}{\uhat}-\logar{2}{\shat}\right)
\right]
.
\eeq
Here we used the shorthand  $\logar{k}{\rhat}=\log^k(|\rhat|/ M_W^2)$
for the logarithms and 
$\cew_{q_\la}=Y_{q_\la}^2/(4\cw^2)+C_{q_\la}/\sw^2$
are the eigenvalues of the electroweak Casimir operator 
for quarks, with $C_{q_\rL}=3/4$  and $C_{q_\rR}=0$. 
Only weak corrections are included in \refeq{oneloopresult}.
This result is obtained in the $M_Z=M_W$ approximation and 
is consistent with the leading- and next-to-leading logarithmic part 
of the one-loop asymptotic expressions%
\footnote{
The correspondence between  \refeq{oneloopresult} and the one-loop expressions 
of \refse{se:corr} is easily seen by means of the relation 
\beq
\cew_{q_\la}-Q^2_{q}=\sum_{V=\mathrm{Z,W^{\pm}}} \left(I^V I^{\bar{V}}\right)_{q_{\lambda}}.
\eeq
}
presented in   \refse{se:corr}.
At two loops, for the complete electroweak NLL corrections%
\footnote{
At the two-loop level the purely weak corrections cannot be isolated 
from the complete electroweak corrections in a gauge-invariant way.
One must thus consider the combination of weak and electromagnetic virtual 
corrections. The latter are regularized by means of a fictitious photon mass 
$\la=\MW$.  This approach is discussed in detail in \citere{Kuhn:2004em}.
}
we obtain
\beqar\label{twolooplogs}
A^{(2)}&\NLLa&
\sum_{\la=\rL,\rR}\Biggl\{
\frac{1}{2}
\left(
Q_{q}\cew_{q_\la}
+
\frac{1}{\sw^2}T^3_{q_\la}
\right)
\Biggl[
Q_{q}\cew_{q_\la}\left(\logar{4}{\shat}-6\logar{3}{\shat}\right)
\nl&&{}
+
\frac{1}{\sw^2}T^3_{q_\la}
\left(\logar{4}{\that}+\logar{4}{\uhat}-\logar{4}{\shat}\right) 
\Biggr]
+\frac{T^3_{q_\la}Y_{q_\la}}{8\sw^4}
\left(\logar{4}{\that}+\logar{4}{\uhat}-\logar{4}{\shat}\right) 
\nl&&{}
+\frac{1}{6}Q_{q}
\Biggl[Q_{q}\left(
\frac{b_1}{\cw^2}\left(\frac{Y_{q_\la}}{2}\right)^2
+\frac{b_2}{\sw^2} C_{q_\la}
\right)
+\frac{1}{\sw^2}T^3_{q_\la}b_2
\Biggr]\logar{3}{\shat}
\Biggr\},
\eeqar
where $b_1=-41/(6\cw^2)$ and $b_2=19/(6\sw^2)$ are the one-loop $\beta$-function coefficients associated with the U(1) and SU(2) couplings, respectively.
Eq.~\refeq{twolooplogs} has been derived from the general 
results  of \citeres{Denner:2003wi,Melles:2001gw}
for leading- and next-to-leading electroweak two-loop logarithms.

\section{Numerical results}
\label{se:numerics}

In this section, we present numerical predictions for 
photon production at high transverse momenta, 
both at the partonic and hadronic level.  
We also compare the cross sections for hadronic photon and $Z$-boson production
and study the impact of the corrections on their ratio.
The lowest order (LO) and 
the next-to-leading-order (NLO) predictions 
for photon production
result from \refeq{generalamplitude}
and  \refeq{generalresult}, respectively.
The next-to-leading-logarithmic%
\footnote{
At the  NLL level, 
angular-dependent logarithms 
are treated as in \citere{Kuhn:2004em}.
} 
(NLL) approximation, at one loop,  corresponds
to the contributions \refeq{bornresult} and \refeq{oneloopresult}.
The next-to-next-to-leading-logarithmic (NNLL) 
predictions, also at one loop, are based on 
the asymptotic expressions \refeq{nnllstracture}--\refeq{heres1b}
combined with the exact result \refeq{Zwfcts} for the counterterm.
Our best predictions, to next-to-next-to-leading-order (NNLO),
include the exact NLO contributions combined with the 
leading and next-to-leading two-loop terms \refeq{twolooplogs}.

The hadronic cross sections are obtained using LO MRST2001 parton
distribution functions (PDFs) \cite{Martin:2002dr}.
We choose $\pT^2$ as the factorization scale
and, similarly as the scale at which the running strong coupling constant
is evaluated. We also adopt the value 
$\al_\rS(M_Z^2)=0.13$ and use the one-loop running expression for 
$\al_\rS(\mu^2)$, in accordance with the LO 
PDF extraction method
of the MRST collaboration. 
We use the
following values of parameters~\cite{Eidelman:2004wy}:  $\alpha=1/137.0$,
$M_Z=91.19 \GeV$, $M_W=80.39 \GeV$ and  
$\sw^2=1-\cw^2=1-M_W^2/M_Z^2$.

\begin{figure}[h]
\vspace*{2mm}
  \begin{center}
\epsfig{file=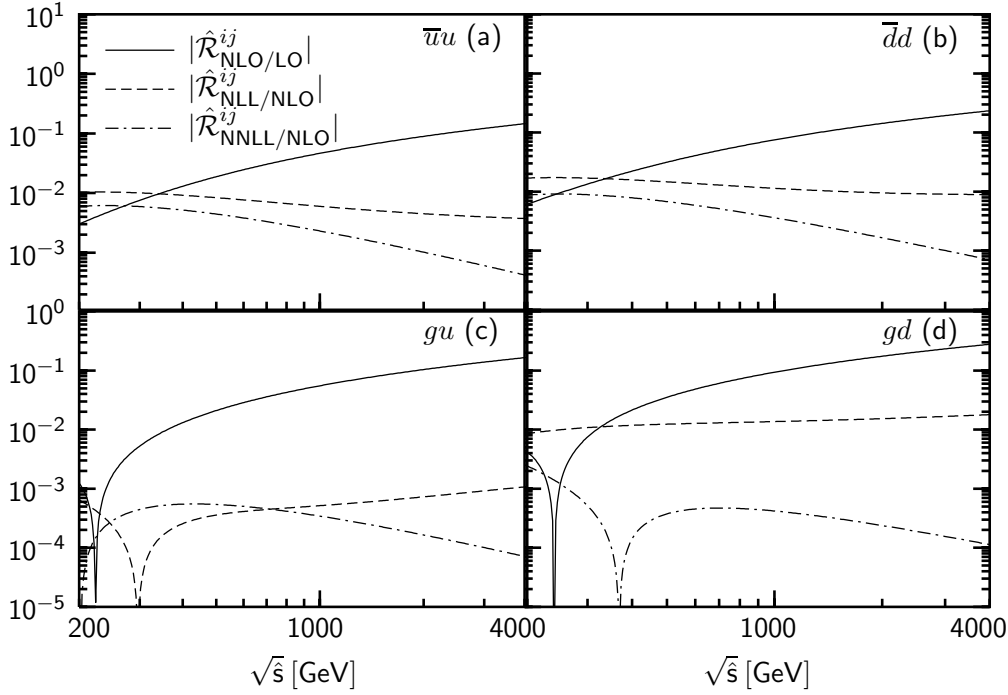, angle=0, width=13cm}
\end{center}
\vspace*{-2mm}
\caption{
Relative one-loop corrections to the partonic differential cross
sections $\rd \hat\si^{ij} / \rd \cos \theta$ at $\cos \theta =0$ 
for (a) $\bar u u$ channel,
(b) $ \bar d d$ channel, (c) $g u$ channel, (d) $ g d$ channel. The
solid, dashed and dot-dashed lines denote the 
modulus of the $\hat {\cal R}$ ratios, 
as defined in the text, for the full NLO cross section, the NLL
approximation and the NNLL approximation of the one-loop cross section, respectively. 
}
\label{fig:part}
\end{figure}

We begin by investigating NLO, NLL and NNLL relative corrections to
the partonic (unpolarized) differential cross section
${\rd \hat{\si}^{i j}}/{\rd \cos\theta}$ 
[see \refeq{partoniccs0}]. 
To this end we define 
\beq
\hat{{\cal R}}^{ij}_{\NLO/\LO}= 
{\rd \hat \si^{ij}_{\NLO}/\rd \cos\theta \over
 \rd \hat \si^{ij}_{\LO} /\rd \cos\theta} 
-1 
\eeq
and similarly ${\hat{\cal R}}^{ij}_{\NLL/\NLO}$ and 
$\hat{\cal R}^{ij}_{\NNLL/\NLO}$.   
These ratios, calculated at  $\cos \theta =0$, are displayed as a
function of $\sqrt \shat$ in \reffi{fig:part}. We consider four
processes: $\bar u u \rar \gamma g$
(\reffi{fig:part}a), $\bar d d \rar \gamma g$ (\reffi{fig:part}b), $g
u \rar \gamma u$ (\reffi{fig:part}c), $g d \rar \gamma d$
(\reffi{fig:part}d). 
The NLO corrections are negative, their absolute size growing
with the partonic energy $\sqrt{\shat}$. 
At $\sqrt{\shat}=4\TeV$ the corrections 
reach $-15\%$ for the $\bar u u$ and $g u$
channels, $-23\%$ for the $\bar d d$, and $-28\%$ for the $g d$ channel.
The NLO correction is well approximated by the NLL result: for 
$\sqrt{\shat}\ge 200\GeV$ the approximation is accurate up to 2\% in all
channels. Even better an approximation of the one-loop weak correction
is provided by the NNLL result. In $\bar u u $ and $\bar d d$ channels the NNLL approximation
differs from the NLO result by less 
than 1\% in the entire region under consideration, whereas  
for the $gu$ and $gd$ channels the quality of the approximation is at
the level of (or better than) one permille.
The absolute LO and NLO cross section 
and the $\hat{\cal R}^{ij}$ ratios 
for specific values of $\cos \theta$, $\sqrt \shat$ and different
collision channels are listed in Table~\ref{tab:part}.

\begin{figure}[]
\vspace*{2mm}
  \begin{center}
\epsfig{file=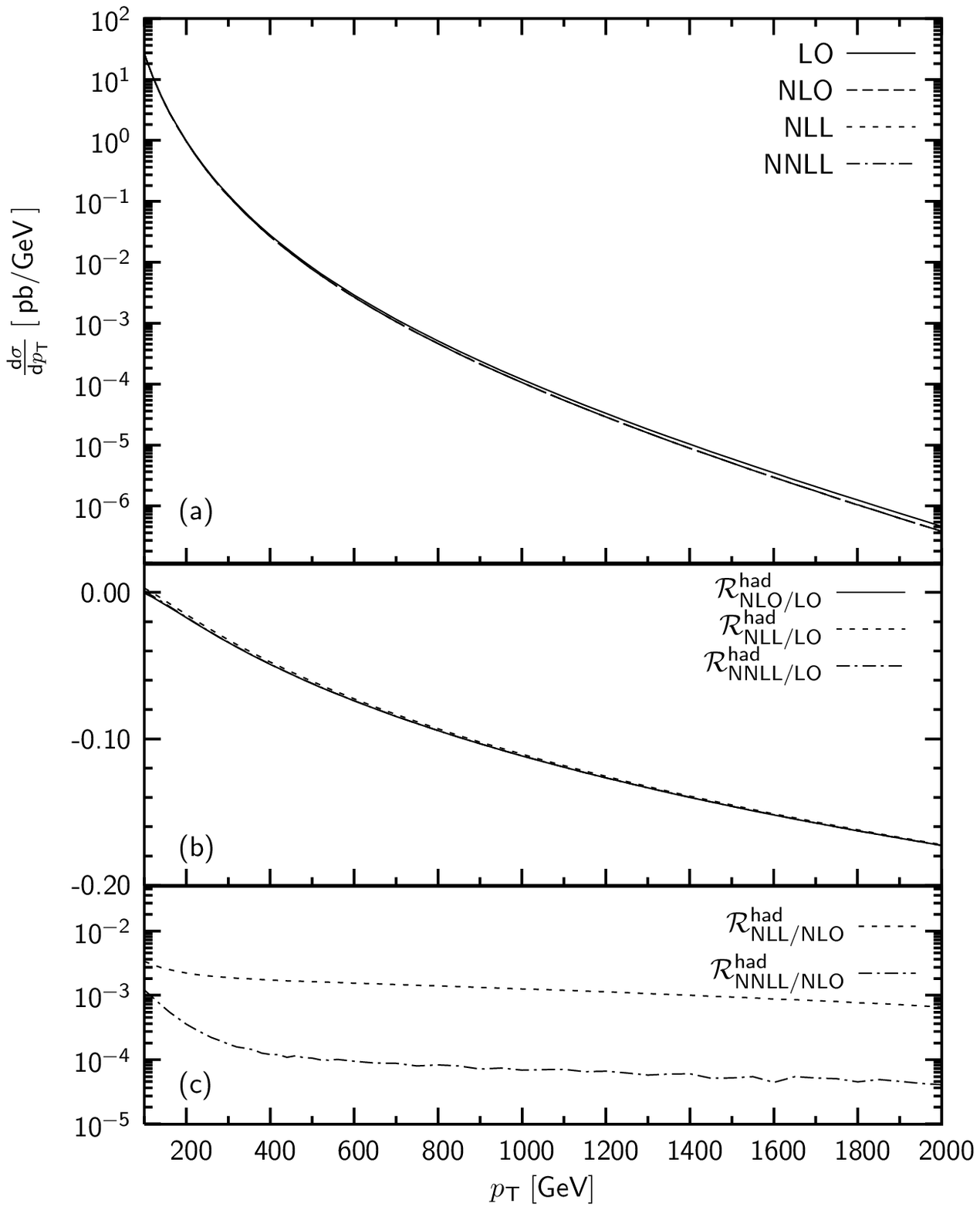, angle=0, width=11.5cm}
\end{center}
\vspace*{-2mm}
\caption{Transverse momentum distribution for $pp\rar \gamma j$ at
  $\sqrt{s}=14 \TeV$.
(a) LO (solid), NLO (dashed),  NLL (dotted) and 
NNLL (dot-dashed) predictions. 
(b) Relative NLO (solid), NLL (dotted) and NNLL (dot-dashed)
weak correction \wrt the LO distribution.
(c) NLL (dotted) and NNLL (dot-dashed) approximations compared to the 
full NLO result.
}
\label{fig:lhc}
\end{figure}

\begin{figure}[]
\vspace*{2mm}
  \begin{center}
\epsfig{file=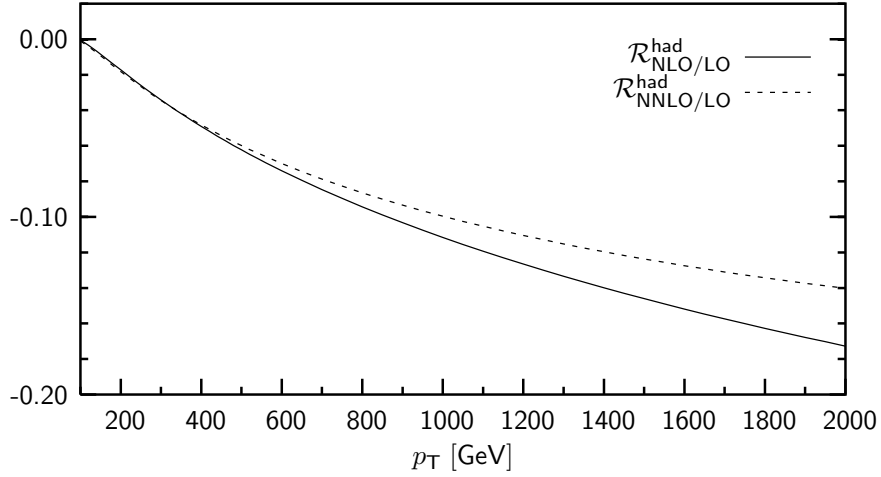, angle=0, width=11.5cm}
\end{center}
\vspace*{-2mm}
\caption{
Relative NLO (solid) and NNLO (dotted) corrections 
to the $p_\rT$ distribution for $pp\rar \gamma j$ at $\sqrt{s}=14 \TeV$.
}
\label{fig:lhcpt12}
\end{figure}

The  transverse momentum distribution $ \rd \si /
\rd\pT$ at the LHC is shown in \reffi{fig:lhc}. 
We display separately the
absolute values of the LO, NLO, NLL
and NNLL differential cross sections (\reffi{fig:lhc}a) and the relative
correction \wrt the LO result for the NLO, NLL and NNLL
distributions (\reffi{fig:lhc}b). The relative correction  \wrt
the LO is now defined as 
\beq
{{\cal R}}^{\rm had}_{\NLO/\LO}= 
{\rd  \si_{\NLO}/\rd \pT \over
 \rd  \si_{\LO} /\rd \pT} 
-1  
\eeq
for the NLO case, and similarly for the NLL and the NNLL cross sections. 
The quality of the NLL and NNLL high-energy  approximations 
of the NLO result is shown in more detail in \reffi{fig:lhc}c. 
The contribution provided by the NLO correction is negative
and increases in size with $\pT$. It ranges from $-6\%$ at 
\mbox{$\pT=500 \GeV$} up to $-17\%$ at $\pT=2 \TeV$.
From Fig.~\ref{fig:lhc}b and Fig.~\ref{fig:lhc}c we conclude that 
the NLL approximation works well, differing from the full NLO prediction by about 
3 permille at low $\pT$ and by less than 1 permille at $\pT\sim 2 \TeV$.
The quality of the NNLL approximation is very good,
at the level of  accuracy of $10^{-3}$ or better in the entire $\pT$ range.
In \reffi{fig:lhcpt12} 
we show the relative size of the corrections 
in the NLO approximation (${{\cal R}}^{\rm had}_{\NLO/\LO}$, solid line), 
and in the approximation which includes the 
next-to-leading logarithmic two-loop terms
 (${{\cal R}}^{\rm had}_{\NNLO/\LO}$, dotted line).
These additional two-loop terms are positive, their size increasing
with $\pT$. At $\pT=2$ TeV they amount to 3\%, yielding the 
total (\ie together with the NLO) correction to the LO of $-14\%$. 

\begin{figure}[]
\vspace*{2mm}
  \begin{center}
\epsfig{file=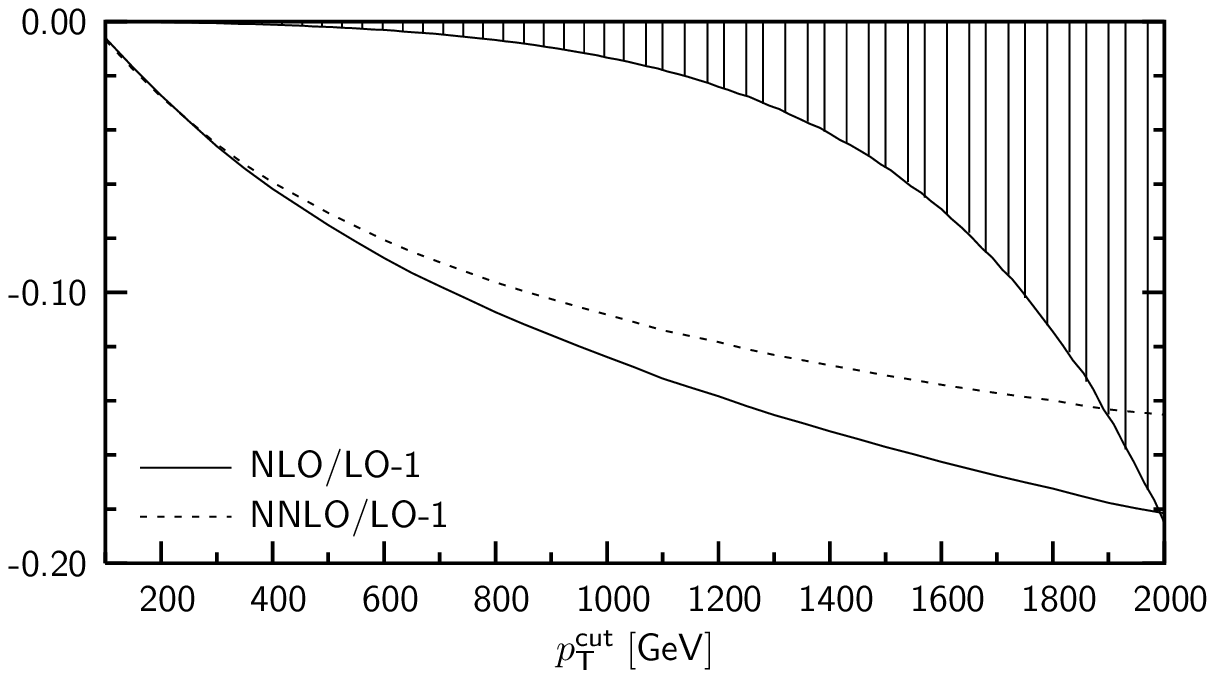, angle=0, width=11.5cm}
\end{center}
\vspace*{-2mm}
\caption{
Relative NLO (solid) and NNLO (dotted) corrections 
\wrt the LO prediction  and statistical
  error (shaded area) for the unpolarized integrated cross section for $pp\rar \gamma j$ at $\sqrt{s}=14 \TeV$ as a function of $\pTcut$.
}
\label{fig:lhctot}
\end{figure}

To underline the relevance of these effects, 
in \reffi{fig:lhctot} we present 
the relative NLO and NNLO 
corrections for the cross section,
integrated over $\pT$ starting from $\pT = \pTcut$, as a function of
$\pTcut$. 
This is compared with the statistical error, defined as
$\Delta \si_{\rm stat} / \si = 1 /\sqrt N$ with $N= \calL \times \si_{\rm LO}$. We assume a total integrated
luminosity $\calL =300 \fba^{-1}$ for the LHC~\cite{LHClum}. 
It is clear from
\reffi{fig:lhctot}, that the size of the one-loop (two-loop) corrections 
is much bigger than (comparable to) the statistical error.
\begin{table}[]
\begin{tabular}{|c|c|c|l|l|r|r|r|}
\hline 
\multicolumn{3}{|c|}{} 
&\multicolumn{2}{|c|}{$\rd\hat\si^{ij} / \rd \cos \theta$ [pb] } 
&
\multicolumn{3}{|c|}{ $\hat{{\cal R}}_{\scriptscriptstyle{\mathrm{A}/\mathrm{B}}}^{ij}\times 10^3$} \\ \hline
$ij$
&$\frac{\sqrt{\hat{s}}}{\GeV}$ 
& $\cos \theta$ 
& \multicolumn{1}{|c|}{$\scriptstyle{\LO}$ }
& \multicolumn{1}{|c|}{$\scriptstyle{\NLO}$ }
& \multicolumn{1}{|c|}{ $\scriptscriptstyle{\NLO/\LO}$}
& \multicolumn{1}{|c|}{$\scriptscriptstyle{\NLL/\NLO}$}
& \multicolumn{1}{|c|}{$\scriptscriptstyle{\NNLL/\NLO}$}
\\ \hline
\hline
$\bar u u$
&   500 &  0.0 & $ 1.5813 \times 10^{  0}$ & $ 1.5516 \times 10^{  0}$ & $-18.819     $ & $8.3754     $ & $4.4462     $\\\      
&  1000 &  0.0 & $ 3.6110 \times 10^{- 1}$ & $ 3.4430 \times 10^{- 1}$ & $-46.526     $ & $5.9974     $ & $2.3120     $\\\      
&  2000 &  0.0 & $ 8.3083 \times 10^{- 2}$ & $ 7.5645 \times 10^{- 2}$ & $-89.517     $ & $4.5064     $ & $1.0219     $\\\hline 
$\bar u u$ 
&   500 &  0.5 & $ 2.6884 \times 10^{  0}$ & $ 2.6425 \times 10^{  0}$ & $-17.068     $ & $10.762     $ & $3.7343     $\\\      
&  1000 &  0.5 & $ 6.1285 \times 10^{- 1}$ & $ 5.8616 \times 10^{- 1}$ & $-43.560     $ & $8.8071     $ & $1.9408     $\\\      
&  2000 &  0.5 & $ 1.4080 \times 10^{- 1}$ & $ 1.2881 \times 10^{- 1}$ & $-85.188     $ & $7.6569     $ & $0.8512     $\\\hline      
\hline 
$\bar d d$ 
&   500 &  0.0 & $ 3.9533 \times 10^{- 1}$ & $ 3.8291 \times 10^{- 1}$ & $-31.427     $ & $14.646     $ & $6.7828     $\\\      
&  1000 &  0.0 & $ 9.0276 \times 10^{- 2}$ & $ 8.3494 \times 10^{- 2}$ & $-75.121     $ & $11.381     $ & $3.6201     $\\\      
&  2000 &  0.0 & $ 2.0771 \times 10^{- 2}$ & $ 1.7810 \times 10^{- 2}$ & $-142.55     $ & $9.5167     $ & $1.6553     $\\\hline   
$\bar d d$ 
&   500 &  0.5 & $ 6.7211 \times 10^{- 1}$ & $ 6.5344 \times 10^{- 1}$ & $-27.769     $ & $19.286     $ & $5.6758     $\\\      
&  1000 &  0.5 & $ 1.5321 \times 10^{- 1}$ & $ 1.4266 \times 10^{- 1}$ & $-68.882     $ & $16.786     $ & $3.0445     $\\\      
&  2000 &  0.5 & $ 3.5200 \times 10^{- 2}$ & $ 3.0501 \times 10^{- 2}$ & $-133.48     $ & $15.642     $ & $1.3809     $\\\hline    
\hline 
$ g u$ 
&   500 &  0.0 & $ 7.4125 \times 10^{- 1}$ & $ 7.2569 \times 10^{- 1}$ & $-20.985     $ & $-0.3584     $ & $-0.5405     $\\\      
&  1000 &  0.0 & $ 1.6927 \times 10^{- 1}$ & $ 1.5994 \times 10^{- 1}$ & $-55.125     $ & $-0.5161     $ & $-0.3435     $\\\      
&  2000 &  0.0 & $ 3.8945 \times 10^{- 2}$ & $ 3.4909 \times 10^{- 2}$ & $-103.63     $ & $-0.7276     $ & $-0.1649     $\\\hline    
$ g u$ 
&   500 &  0.5 & $ 6.3010 \times 10^{- 1}$ & $ 6.1387 \times 10^{- 1}$ & $-25.756     $ & $-2.2604     $ & $-0.2459     $\\\      
&  1000 &  0.5 & $ 1.4364 \times 10^{- 1}$ & $ 1.3469 \times 10^{- 1}$ & $-62.295     $ & $-2.6157     $ & $-0.1587     $\\\      
&  2000 &  0.5 & $ 3.3000 \times 10^{- 2}$ & $ 2.9263 \times 10^{- 2}$ & $-113.22     $ & $-3.0577     $ & $-0.0747     $\\\hline     
\hline 
$ g d$ 
&   500 &  0.0 & $ 1.8531 \times 10^{- 1}$ & $ 1.7880 \times 10^{- 1}$ & $-35.167     $ & $-12.356     $ & $-0.3746     $\\\      
&  1000 &  0.0 & $ 4.2317 \times 10^{- 2}$ & $ 3.8365 \times 10^{- 2}$ & $-93.383     $ & $-13.653     $ & $-0.4196     $\\\      
&  2000 &  0.0 & $ 9.7363 \times 10^{- 3}$ & $ 8.0397 \times 10^{- 3}$ & $-174.24     $ & $-15.298     $ & $-0.2382     $\\\hline     
$ g d$ 
&   500 &  0.5 & $ 1.5753 \times 10^{- 1}$ & $ 1.4944 \times 10^{- 1}$ & $-51.298     $ & $-18.358     $ & $-0.0453     $\\\      
&  1000 &  0.5 & $ 3.5909 \times 10^{- 2}$ & $ 3.1686 \times 10^{- 2}$ & $-117.62     $ & $-20.363     $ & $-0.1781     $\\\      
&  2000 &  0.5 & $ 8.2500 \times 10^{- 3}$ & $ 6.5458 \times 10^{- 3}$ & $-206.57     $ & $-23.109     $ & $-0.1086     $\\\hline     
\end{tabular}
\caption{
Absolute value of the LO and NLO  partonic 
differential cross
section $\rd\hat\si^{ij} / \rd \cos \theta$ 
and the ratios  
$\hat{{\cal R}}^{ij}_{\NLO/\LO}$, $\hat{{\cal R}}^{ij}_{\NLL/\NLO}$ and 
$\hat{{\cal R}}^{ij}_{\NNLL/\NLO}$ in permille,
for the  partonic processes 
$\bar uu\to \gamma g$, 
$\bar dd\to \gamma g$, 
$gu\to \gamma u$ and
$gd\to \gamma d$.
The running strong coupling $\alpha_\rS(\mu^2)$
is taken at the scale $\mu^2=p_\rT^2=
(1-\cos^2\theta)\shat/4$.
}
\label{tab:part}
\end{table}

\pagebreak

\begin{figure}[]
\vspace*{2mm}
  \begin{center}
\epsfig{file=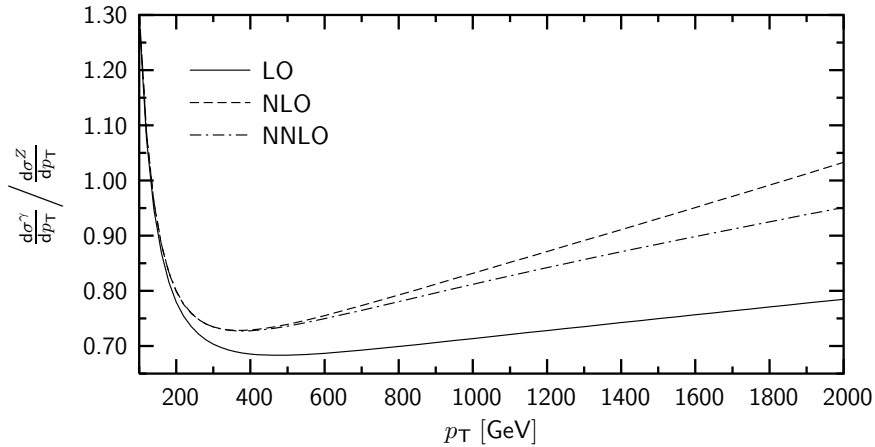, angle=0, width=11.5cm}
\end{center}
\vspace*{-2mm}
\caption{Ratio of the transverse momentum distributions
for the processes $pp\rar \gamma j$ and  $pp\rar Z j$ 
at $\sqrt{s}=14 \TeV$: 
LO (solid), NLO(dashed) and NNLO (dot-dashed) predictions.
}
\label{fig:zgammaratio}
\end{figure}

In \reffi{fig:zgammaratio}, we plot the ratio of the 
$\pT$ distributions of photons and $Z$ bosons.
The latter is computed within the \msbar~scheme using the same input 
parameters as in  \citere{Kuhn:2005az}. 
Such ratio is expected to be less sensitive to theoretical errors
than the distributions themselves, since many
uncertainties such as the scale at which
$\alpha_\rS$ is calculated or the choice of PDFs 
cancel to a large extend in the ratio. Moreover, due to a similar
cancellation mechanism, the ratio should remain
stable against QCD corrections.
From \reffi{fig:zgammaratio} we observe that the weak 
corrections modify the production ratio considerably. The effect is
the strongest at high $\pT$. In this region, the LO photon cross
section is smaller than the cross section for $Z$ boson
production by about 25\%. The relatively large NLO corrections for $Z$
production, as compared to $\gamma$ production, cause the full NLO production
rates to become equal at the highest $\pT$ considered here, \ie
$\pT\sim 2$ TeV. The two-loop corrections modify the ratio and lead
to a few percent decrease at
high $\pT$.

\begin{figure}[]
\vspace*{2mm}
  \begin{center}
\epsfig{file=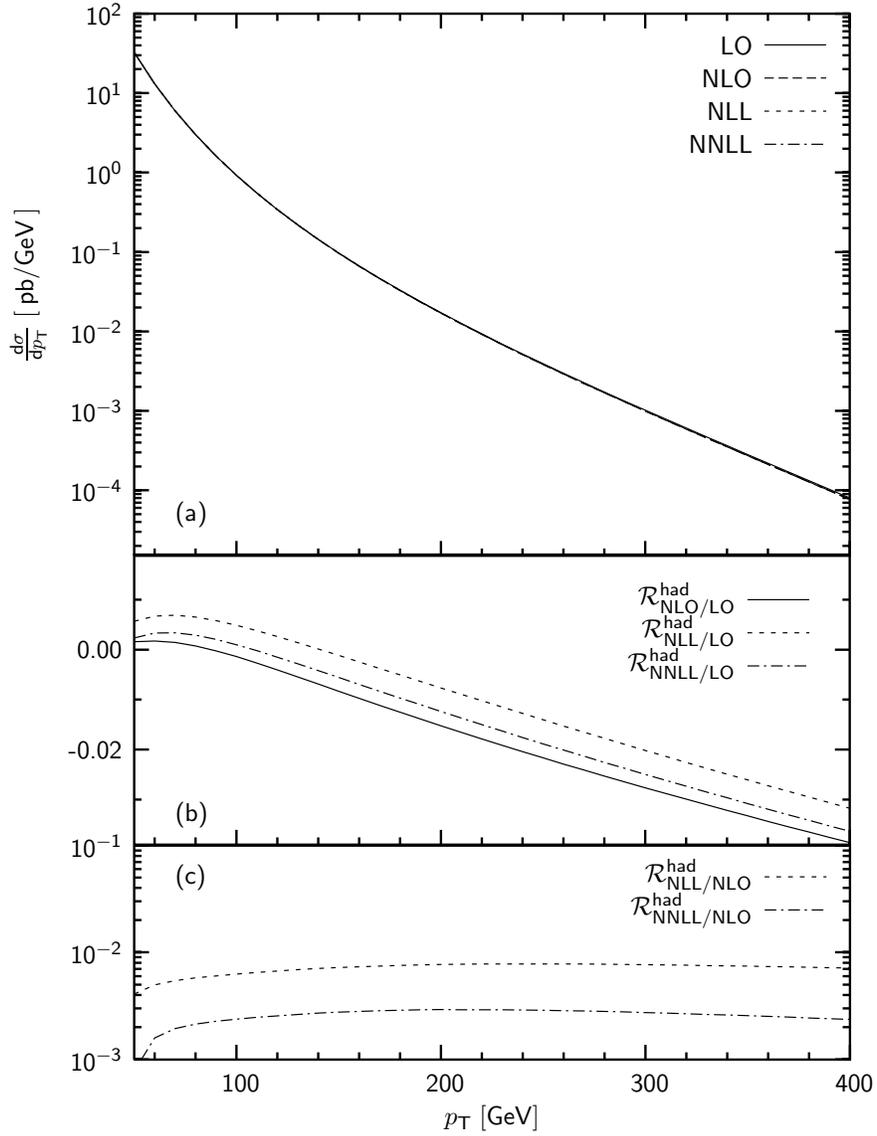, angle=0, width=11.5cm}
\end{center}
\vspace*{-2mm}
\caption{Transverse momentum distribution for $p\bar p\rar \gamma j$ at
  $\sqrt{s}=2 \TeV$.
(a) LO (solid), NLO (dashed),  NLL (dotted) and 
NNLL (dot-dashed) predictions. 
(b) Relative NLO (solid), NLL (dotted) and NNLL (dot-dashed)
weak correction \wrt the LO distribution.
(c) NLL (dotted) and NNLL (dot-dashed) approximations compared to the 
full NLO result.
}
\label{fig:tev}
\end{figure}

\begin{figure}[]
\vspace*{2mm}
  \begin{center}
\epsfig{file=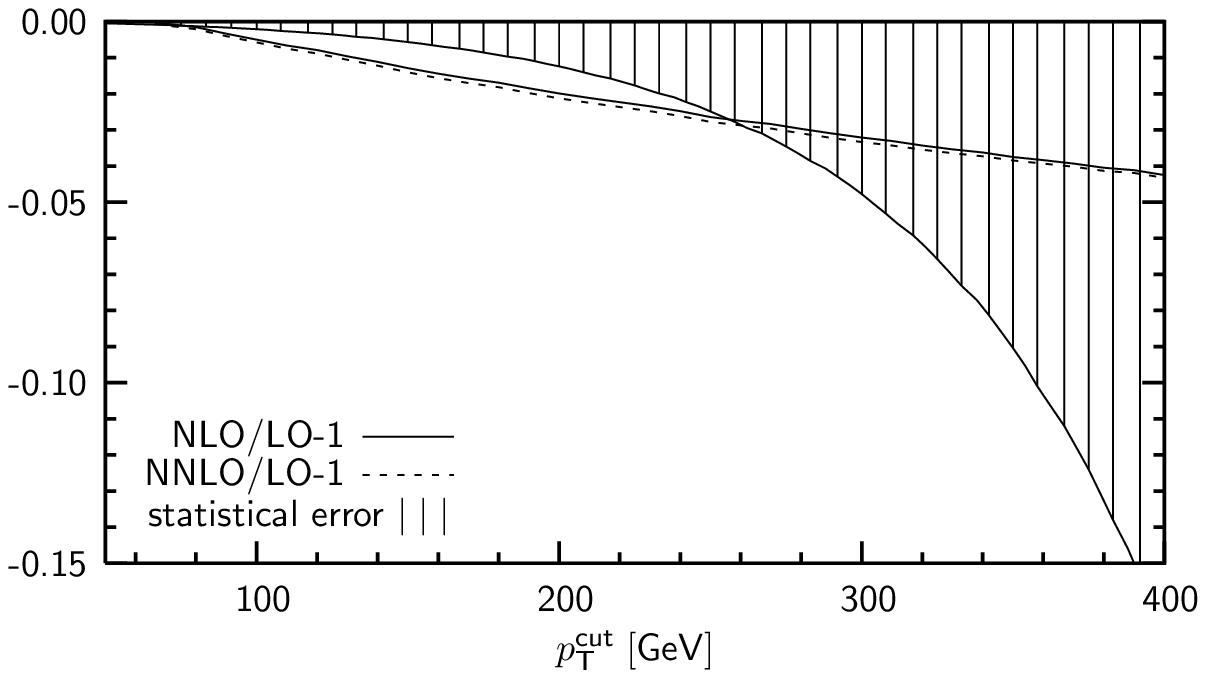, angle=0, width=11.5cm}
\end{center}
\vspace*{-2mm}
\caption{Relative NLO (solid) and NNLO (dotted) corrections \wrt the LO and statistical
  error (shaded area) for the unpolarized integrated cross section for $p\bar p\rar
  \gamma j$ at $\sqrt{s}= 2 \TeV$ as a function of $\pTcut$.
}
\label{fig:tevtot}
\end{figure}

\begin{figure}[]
\vspace*{2mm}
  \begin{center}
\epsfig{file=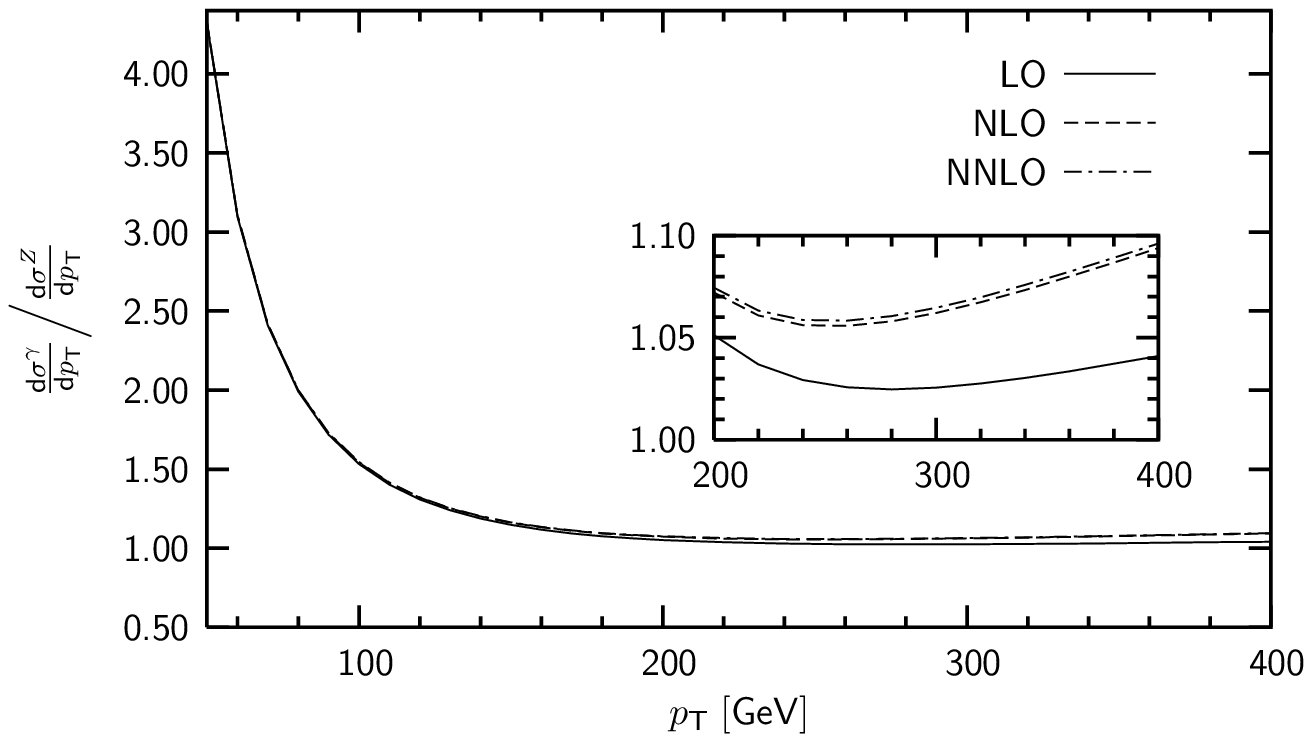, angle=0, width=11.5cm}
\end{center}
\vspace*{-2mm}
\caption{Ratio of the transverse momentum distributions
for the processes $p \bar p\rar \gamma j$ and  $p \bar p\rar Z j$ 
at $\sqrt{s}=2 \TeV$:
LO (solid), NLO(dashed) and NNLO (dot-dashed) predictions.
}
\label{fig:tevzgammaratio}
\end{figure}

We perform a similar analysis for direct
photon production at high transverse momentum at the Tevatron.
In  \reffi{fig:tev} we show the transverse momentum distribution
(\reffi{fig:tev}a), the relative size of the corrections (\reffi{fig:tev}b)
and the quality of the one-loop NLL and NNLL approximations
(\reffi{fig:tev}c).
The effects of weak corrections are generally much smaller for the
case of the Tevatron than the LHC, with the NLO corrections not
exceeding $-4\%$ at the highest $\pT$ considered, \ie at $\pT\sim 400$ GeV.
In \reffi{fig:tevtot} the relative size of the corrections 
to the integrated cross section is compared 
with the statistical error expected for an integrated 
luminosity  $\calL =11 \fba^{-1}$~\cite{TEVlum}.
At the energies of the Fermilab collider,
the NLO weak correction is of the order of the 
statistical error and we conclude it should be taken into account 
when considering precision measurements.
The two-loop terms turn out to be negligible.
The ratio of the 
$\pT$ distributions of photons and $Z$ bosons 
is shown in \reffi{fig:tevzgammaratio}. Since the
weak corrections to the $Z$ and $\gamma$ production at the Tevatron
are moderate, their effect on the ratio is fairly small and stays within a 
few percent range for all values of $\pT$ considered here.

Our numerical results, both for the case of the LHC and the Tevatron,
are in agreement with the results presented in figures of~\citere{Maina:2004rb}.

\section{Summary}
\label{se:conc}

In this work we present one-loop weak corrections to
the direct production of photons with large transverse
momenta at hadron colliders.
Analytical results are given for the parton subprocess $\bar
q q \to  \gamma g$ and its crossed versions. We also present
approximate expressions valid in the  high-energy region, $\shat\gg
M_W^2$, where the weak corrections are enhanced by logarithms of $\shat/M_W^2$.
The complete high-energy approximation discussed in this paper
includes all large logarithms  as well as those terms that are not
logarithmically enhanced.
This approximation is in very good agreement with the complete one-loop result.  We also calculate the two-loop electroweak corrections in
the next-to-leading logarithmic approximation.

The corrections are then evaluated numerically for proton-antiproton
collisions at 2 TeV (Tevatron) 
and proton-proton collisions at 14 TeV (LHC) in the region of large tranverse momentum ($p_\rT$). 
The corrections are negative and their size increases with $p_\rT$. At the Tevatron, transverse momenta up to 400 GeV will be explored and the weak corrections 
may reach up to $-4\%$. At the LHC, transverse momenta of 2 TeV or
more are within the reach. In this region the one-loop corrections are significant,
about $-17\%$, and even the dominant two-loop logarithmic terms
must be included in precise predictions.
Finally we compare the cross sections for photon and $Z$-boson production and find that the
electroweak corrections have a considerable impact on their ratio at the LHC.

\section*{Acknowledgements}
This work was supported in part by 
BMBF Grant No.~05HT4VKA/3
and by the Deutsche Forschungsgemeinschaft 
(Sonderforschungsbereich Transregio SFB/TR-9 
``Computational Particle Physics''). 
M.~S. would like to acknowledge financial 
support from the Graduiertenkolleg "Hochenergiephysik und Teilchenastrophysik".

%

\pagebreak

\end{document}